\pdfoutput=1
\documentclass{article}			
\usepackage[T1]{fontenc}
\usepackage[utf8]{inputenc}
\usepackage{graphicx}   
\usepackage{amsmath,amssymb}
\usepackage{graphicx}
\usepackage{adjustbox}
\usepackage{booktabs}
\usepackage{hyperref}
\usepackage{threeparttable}
\usepackage{authblk}
\usepackage{geometry}
\usepackage{rotating}
\usepackage{booktabs,siunitx,pgfplots,subcaption}
\usepackage{tikz}
\usepackage{quantikz}
\usetikzlibrary{arrows.meta,positioning,calc,decorations.pathreplacing,fit} 
\usepackage{subcaption} 
\usepackage{caption} 
\usepackage{graphicx}
\usepackage{xcolor}
\graphicspath{{./}{./figs/}}
\DeclareGraphicsExtensions{.pdf,.png,.jpg}

\title{Diagram-to-Circuit QNLP for Financial Sentiment Analysis}
\author{Takayuki Sakuma\\
Faculty of Economics, Soka University \\ \text{e-mail: tsakuma@soka.ac.jp}
}
\date{\today}

\begin{document}
\maketitle

\begin{abstract}
We study a \emph{QDisCoCirc}-inspired, chunked diagram-to-circuit quantum natural language processing (QNLP) model for three-class sentiment classification of financial texts. In our classical simulations, we keep the Hilbert-space dimension manageable by decomposing each sentence into short contiguous chunks. Each chunk is mapped to a shallow quantum circuit, and the resulting Bloch vectors are used as a sequence of quantum tokens. Simple averaging of chunk vectors ignores word order and syntactic roles. We therefore add a small Transformer encoder over the raw Bloch-vector sequence and attach a CCG-based type embedding to each chunk. This hybrid design preserves physically interpretable semantic axes of quantum tokens while allowing the classical side to model word order and long-range dependencies. The sequence model improves test macro-F1 over the averaging baseline and chunk-level attribution further shows that evidential mass concentrates on a small number of chunks, that type embeddings are used more reliably for correctly predicted sentences. For real-world quantum language processing applications in finance, future key challenges include circuit designs that avoid chunking and the design of inter-chunk fusion layers.
\end{abstract}

\section{Introduction}
Large language models (LLMs) are increasingly adopted in finance for tasks such as sentiment analysis and risk measurement. For example, FinBERT~\cite{araci2019finbert} fine-tunes a pre-trained BERT on financial news and reports for sentiment and risk classification; BloombergGPT~\cite{wu2023bloomberggpt} is a 70-B parameter GPT style model trained by Bloomberg; and FinGPT~\cite{yang2023fingpt} provides stock-price signal extraction and algorithmic-trading chatbots, along with an open data-augmentation pipeline.

Despite their success, Transformer-based models remain hard to interpret: verifying what they have learned and how they reason is difficult. Although attention-weight visualizations are common, they have been widely criticized~\cite{jain2019attention,wiegreffe2019attention}. Probing methods that analyze hidden states with linear classifiers and related techniques can confound the expressivity of the probe with the model’s knowledge, leaving it unclear whether linguistic structure is encoded by the model or by the probe itself~\cite{belinkov2019analysis,hewitt2019structural}. Moreover, in regulated settings where auditing and accountability are required, post-hoc explanations of black-box systems may be insufficient~\cite{rudin2019stop}.

In response to these challenges, mechanistic interpretability has begun to be applied to financial NLP. Tatsat \emph{et al.}~\cite{tatsat2025beyond} demonstrate tools such as Logit Lens, activation patching, and sparse autoencoders to localize layer- and head-level contributions in financial settings and to intervene on features learned by the model~\cite{tatsat2025beyond}. They argue that, beyond post-hoc rationalizations, causal and intervention-based analyses are necessary to meet regulatory expectations.

This paper investigates how a \emph{QDisCoCirc}-inspired, chunked diagram-to-circuit model enables causal and intervention-based analysis in quantum natural language processing. Combinatory Categorial Grammar (CCG) is a framework that systematically composes word-meaning representations according to grammatical structure. Laakkonen \emph{et al.}~\cite{laakkonen2024algorithms} extended DisCoCirc, which generalizes CCG to text circuits, to quantum computation and proposed efficient algorithms for constructive text processing; they referred to the resulting framework as \emph{QDisCoCirc}. QDisCoCirc performs compositional semantics using sequences of quantum gates and provides explicit interpretability by identifying, gate by gate, which word influences which qubit. Duneau \emph{et al.}~\cite{meichanetzidis2024scalable} implemented QDisCoCirc on Quantinuum’s H1-1 trapped-ion quantum processor. By reusing qubits under the constraints of the 20-qubit H1-1 device, they managed to fit circuits of up to 108 logical qubits into just 20 physical qubits. They confirmed that there was no significant loss of accuracy even for texts containing up to 30 entities (noun phrases), demonstrating the scalability of their approach to longer texts.

In this study, we evaluate a \emph{QDisCoCirc}-inspired, chunked diagram-to-circuit QNLP model on sentiment analysis of financial texts via classical simulation. 
The task is sentiment classification with three classes-negative, neutral, and positive-extending the binary QNLP experiments of Duneau \emph{et al.}~\cite{meichanetzidis2024scalable}.
By applying this \emph{QDisCoCirc}-inspired approach to realistic financial sentiment tasks, we provide a finance-oriented case study and introduce intervention-based explanation metrics that quantify the roles of the quantum and classical components. While Duneau \emph{et al.}~\cite{meichanetzidis2024scalable} discuss inspectability only at the circuit-component (word state/box) level, our study differs in that it allows us to decompose, within a sentence, the attribution of “which information channels contributed and to what extent” into three components: (a) Bloch-vector representations (semantic axes), (b) CCG types (syntactic channel), and (c) type gates (control of throughput in the syntactic channel). 
In this way, our study complements the hardware-based demonstration of scalability by Duneau \emph{et al.}~\cite{meichanetzidis2024scalable} and provides a novel evaluation perspective that enables quantitative tracking of the respective roles of quantum (Bloch) and syntactic (CCG) factors at the sentence level.

Nonetheless, the empirical evaluation in this paper currently relies on classical simulations. Due to computational resource constraints, we adopt a design that divides sentences into multiple chunks to keep the circuit width manageable, embeds each chunk independently into a quantum circuit, and then constructs the sentence representation by combining the resulting density matrices via a convex combination. 
Although this strategy-segmenting a sentence into short chunks and subsequently recombining them via a convex mixture-provides a feasible way to process long texts without incurring exponential circuit width, it may fail to capture pragmatic and syntactic long-range dependencies that span across chunk boundaries, potentially becoming a limiting factor for predictive accuracy.

\paragraph{Terminology and positioning.}
The QDisCoCirc framework \cite{meichanetzidis2024scalable} builds on the DisCoCirc ``text-as-circuit'' view and aims at compiling entire sentences (and, in principle, longer texts) into executable circuits. In this paper, due to classical-simulation constraints, we do not attempt to compile each whole sentence into a single end-to-end text circuit. Instead, we instantiate compositional circuits at the level of short CCG-derived chunks: each chunk is converted into a DisCoCat-style pregroup diagram and mapped to a shallow quantum circuit, while inter-chunk context is handled classically (pooling or a small Transformer). We therefore use the term \emph{QDisCoCirc-inspired} when referring to our instantiation.

Research has also explored the use of Combinatory Categorial Grammar (CCG) in classical LLMs. Tian \emph{et al.} proposed a framework that introduces CCG supertags and uses information obtained during the decoding of these tags to guide attention weights over input words~\cite{tian2023ccg-easa}. Since supertags can assign semantic functions to words, they enable the attention mechanism to capture semantic cues that are difficult to model using conventional dependency structures alone. Additionally, Zhao and Penn proposed a method called \emph{LLM\mbox{-}supertagger}, which uses LLMs for CCG supertagging~\cite{zhao2024llm-supertagger}. Although the present study is similar in spirit to these “CCG-guided attention” approaches, it differs in that it serializes sentences while preserving semantic axes on the quantum side and allows the syntactic roles of attention to be visualized and quantified through type embeddings. 

In recent years, the power consumption and carbon-dioxide emissions associated with LLM training and inference have become major concerns. Strubell \emph{et al.}~\cite{strubell2019energy} estimated that neural architecture search for Transformer models could emit up to \(\,626{,}155\,\mathrm{lb}\) (approximately \(284\,\mathrm{tCO_2e}\)) of CO\(_2\)-equivalent, which is comparable to the lifecycle emissions of five passenger cars. Patterson \emph{et al.}~\cite{patterson2021carbon} reported that training GPT-3 (175B) required a total energy of \(1{,}287\,\mathrm{MWh}\) (1.287 GWh) and resulted in net emissions of \(552.1\,\mathrm{tCO_2e}\). These figures underscore the need for alternative computational paradigms with better long-run energy scalability. Although practical quantum devices are still limited, the theoretical resource requirements of quantum models suggest that they may offer a more energy-efficient path in the long term, motivating an exploration of their integration into language-model pipelines.

\section{Model}
QDisCoCirc is a framework that represents sentences as quantum circuits built from sequential and parallel compositions of small-scale components. Due to the limits of classical simulation, our implementation constructs shallow circuits only at the level of short CCG-derived chunks (via DisCoCat-style pregroup diagrams) and models inter-chunk context on the classical side.
Content words such as nouns are represented as input states (“states”); verbs and adjectives are processes (“boxes”) that transform or combine states; and sentence evaluation is represented as an effect. Sentences can also be connected via circuit composition, and semantic similarity between sentences is evaluated by the overlap of the output states of the corresponding circuits. Grammar serves as a wiring specification that determines which wires are contracted and which remain.

\subsection{Basic Structure of Quantum Circuits}
As an example, we map \emph{“The company increased profits.”} to a quantum circuit. The mapping from sentence to circuit proceeds in the following five steps.

\begin{enumerate}
  \item \textbf{Role Assignment} \;   Using a standard CCG, we assign syntactic roles to the words in the sentence (subject, object, predicate). CCG is a type of formal grammar that assigns a functional category to each word and constructs sentences according to combinatory rules. The predicate \emph{increased} is specified to take the subject (\emph{company}) on the left and the object (\emph{profits}) on the right.

  \item \textbf{Typing (\(n,s\))} \;
  \(\mathrm{company}:n,\;\mathrm{profits}:n,\;
        \mathrm{increased}:(n^r\,s\,n^l)\).
  The \(n^r\) and \(n^l\) flanking the verb are the argument types that connect with the subject and object, respectively. Here \textbf{\(n\)} denotes the noun type and \textbf{\(s\)} denotes the sentence type. The superscripts \(n^r\) and \(n^l\) can be regarded simply as symbols indicating partners that will be eliminated in the subject and object positions.
  
  \item \textbf{Diagrammatic Construction (Wiring Template)} \;
  We connect \(n\) with \(n^r\) and \(n^l\) with \(n\) using cups, and then contract these pairs, leaving a single \(\,s\)-wire as the core of the sentence. “Cups” and “caps” are wiring templates that connect lines of complementary types to eliminate them. In computational terms, this corresponds to taking an \textit{inner product} (contraction), leaving only the skeleton of the words and the sentence. A cup connects and eliminates adjacent types and their adjoints (e.g., \(n\) and \(n^r\)), contracting the wiring; in this example it eliminates the \(n\text{--}n^r\) and \(n^l\text{--}n\) pairs and leaves \(s\). By contrast, a cap generates two wires and is used when constructing label density matrices or traces. Cups and caps are fundamental building blocks for eliminating (cups) or copying (caps) words via wiring.
  
  \item \textbf{Quantum Circuitization (Figure~1)} \;
  We assign qubits to each type. In this implementation, we allocate one qubit to \(n\) and typically one qubit to \(s\); if the circuit would otherwise be empty, we synthesize \(\mathrm{I}\!\to\! s\) to ensure the presence of an \(s\)-wire. Each word is represented by a parametric quantum circuit (single-qubit rotations plus entangling gates). The circuit parameters for \emph{increased}, such as \((\theta,\phi)\), contribute to the state of the \(s\)-wire responsible for sentiment via the diagrammatic mapping and the IQP ansatz. 

\item \textbf{Classification} \;
The sentence is syntactically parsed using CCG and automatically split into chunks (phrases) that preserve word order. To prevent an explosion in the number of qubits and in circuit depth, each chunk is limited to a maximum length of \(5\) words (chunks that are too long are further subdivided). Chunks whose types reduce to the identity \(I\) (where \(I\otimes X \cong X\)), the unit of the tensor product, correspond to empty circuits: they have no outputs to be measured, so no Bloch vector is obtained in the final measurement. Such chunks are therefore excluded, and their contribution to the sentence representation is set to \(0\).

\begin{figure}[t]
  \centering
  \begin{quantikz}[row sep={0.6cm,between origins}, column sep=0.45cm]
    \lstick{$\ket{0}^{n}$ (company)}   & \gate{R_Z(\alpha_1)} & \gate{R_X(\beta_1)} & \qw \rstick{\scriptsize $n$} \\
    \lstick{$\ket{0}^{s}$ (increased)} & \gate{R_Y(\theta)}   & \gate{R_Z(\phi)}    & \qw \rstick{\scriptsize $n^r\; s\; n^l$} \\
    \lstick{$\ket{0}^{n}$ (profits)}   & \gate{R_Z(\alpha_2)} & \gate{R_X(\beta_2)} & \qw \rstick{\scriptsize $n$}
  \end{quantikz}
  \caption{Quantum circuit for \emph{“The company increased profits.”} Each chunk (a semantically meaningful unit such as the subject, verb, or object) is prepared as single qubit. In the type notation, the subject is denoted by $n$, the verb by $n^r s n^l$, and the object by $n$. The superscript $n$ in $\ket{0}^{n}$ indicates the noun type in DisCoCat: $n$ denotes the noun (subject/object) type, $n^r$ and $n^l$ denote the right and left adjoints associated with the verb (indicating that they can contract with the subject and object), and $s$ denotes the sentence type. In this example, the grammatical roles “subject”, “verb”, and “object” coincide with the chunks, with one qubit assigned to each. In general, however, one qubit is assigned to a single chunk, and a chunk may contain multiple words that are grouped together during preprocessing (e.g., complex prepositional phrases, relative clauses, etc.).}
  \label{fig:example_circuit}
\end{figure}

\medskip

\noindent
Let the set of valid chunks be
\[
  K = \{\, j \mid U_j \text{ has a CCG type that yields an } s\text{-typed output}\,\},
\]
and denote its cardinality by \(|K| = M\). For each valid chunk \(j \in K\), let
\(r_j = (r_{xj}, r_{yj}, r_{zj})\) be the corresponding Bloch vector, and lift it to a
density matrix
\[
\rho(r_j)=\tfrac12\!\left(I+r_{xj}\sigma_x+r_{yj}\sigma_y+r_{zj}\sigma_z\right).
\]
The sentence representation is then defined as the uniform convex combination
\[
  \rho_{\text{doc}}=\frac{1}{M}\sum_{j\in K}\rho(r_j).
\]
(If \(M=0\), the sentence in question is skipped during training.)
While this simple mean is easy to implement and robust, syntactic cues such as
word order and long-range dependencies are abstracted away, so we also evaluate a separate model.

We compute the overlap between the sentence representation obtained by chunk
aggregation, \(\rho_{\text{doc}}\in\mathcal{D}(\mathbb{C}^2)\), and the label density
matrices corresponding to each class \(c\in\{0,1,2\}\) (Neg/Neu/Pos), denoted by
\(\{\sigma_{c,j}\}_{j=1}^{K_c}\subset\mathcal{D}(\mathbb{C}^2)\). The label density
matrices for each class \(c\) are stored as learnable Bloch-vector parameters \(\{\,r_{c,j}\in\mathbb{R}^3\,\}_{j=1}^{K_c}\), which are converted during
evaluation via \(\sigma_{c,j}=\rho(r_{c,j})\). The similarity between
\(\rho_{\text{doc}}\) and \(\sigma_{c,j}\) is measured using the normalized
Hilbert--Schmidt inner product:
\[
  s(\rho_{\text{doc}},\sigma_{c,j})
   \;=\;
  \frac{\operatorname{tr}(\rho_{\text{doc}}\,\sigma_{c,j})}
       {\sqrt{\operatorname{tr}(\rho_{\text{doc}}^2)\,\operatorname{tr}(\sigma_{c,j}^2)}}
   \in [0,1].
\]
For the case where there are multiple label density matrices (\(K_c>1\)), we
aggregate them using a temperature-scaled log-sum-exp:
\[
  S_c \;=\;
  m_c + \tau\log\!\sum_{j}\exp\!\bigl(\tfrac{s_{c,j}-m_c}{\tau}\bigr)
  - \tau \log K_c,
\]
\end{enumerate}
where \(s_{c,j}=s(\rho_{\text{doc}},\sigma_{c,j})\), \(m_c=\max_j s_{c,j}\), and
\(\tau>0\). In this paper, we set \(\tau=0.1\) and \(K_c=3\). As \(\tau\!\to\!0\), this
expression converges to \(\max_j s_{c,j}\). Finally, the class scores \(\{S_c\}\)
for an input sentence \(x\) are transformed into probabilities
\(p(y=c\mid x) = \mathrm{softmax}_c(S_c)\), which are then fed into the
cross-entropy loss function for training.

For each class \(c\), we construct the label density matrices as follows. We first gather the density matrices of all chunks from sentences with label \(c\), and then sequentially select representatives that are as far apart from each
other as possible. Specifically, we choose as the reference the density matrix whose Bloch vector has the largest norm. Subsequent representatives are chosen by repeatedly adding the matrix that is most distant---according to a dissimilarity measure based on the Hilbert--Schmidt inner product---from those already selected. This so-called farthest-first rule yields the three representatives \(\sigma_{c,1},\sigma_{c,2},\sigma_{c,3}\).

\subsection{Normalized Similarity and Threshold Optimization}
During training, we use the Hilbert--Schmidt inner product because its gradient and Hessian are easier to handle and more stable for backpropagation. However, at evaluation, to emphasize directional differences between prototypes and to suppress biases caused by differences in norms (scales), we evaluate sentence vectors and label prototypes using cosine similarity:
\[
  s_k(x)=\cos\!\bigl(\hat{w}_k,\hat{h}(x)\bigr),
  \qquad
  \hat{v}=\frac{v}{\lVert v\rVert}.
\]
Specifically, a density matrix $\rho$ is mapped to its Bloch representation vector $\mathbf{b}(\rho)$ and normalized as
\[
  \widehat{\mathbf{b}}(\rho)\;=\; \frac{\mathbf{b}(\rho)}{\lVert \mathbf{b}(\rho)\rVert_2}.
\]
This normalized vector is then used in the cosine similarity.

Additionally, after training is completed, class-specific threshold optimization is applied once as a final calibration step. Specifically, for the dev (validation) set $\mathcal{D}_{\mathrm{dev}}$, the thresholds $\boldsymbol{\tau}=(\tau_1,\dots,\tau_C)$ are tuned via grid search so that
\begin{equation}
 \mathrm{F1}_{\mathrm{macro}}(\boldsymbol{\tau})
  =\frac{1}{C}\sum_{k=1}^{C}\mathrm{F1}_k(\boldsymbol{\tau}),
\qquad
\mathrm{F1}_k(\boldsymbol{\tau})
=\frac{2\,\mathrm{TP}_k(\boldsymbol{\tau})}{2\,\mathrm{TP}_k(\boldsymbol{\tau})+\mathrm{FP}_k(\boldsymbol{\tau})+\mathrm{FN}_k(\boldsymbol{\tau})},
  \label{eq:macro-f1}
\end{equation}
is maximized, and the $\boldsymbol{\tau}^\star$ that achieves this maximum is adopted. At inference time, this $\boldsymbol{\tau}^{\star}$ is used for the probability outputs $p_k(x)$, and the predicted label $\hat{y}_{\boldsymbol{\tau}}(x)$ is determined as
\[
  \hat{y}_{\boldsymbol{\tau}}(x)=
  \begin{cases}
    \arg\max_{k\in\mathcal{C}(x)} p_k(x), & \mathcal{C}(x)\neq\varnothing,\\[3pt]
    \arg\max_k p_k(x), & \text{otherwise},
  \end{cases}
\]
where $\mathcal{C}(x)=\{k\mid p_k(x)\ge \tau_k\}$.

\section{Supplementary Processing}
The following supplementary procedures are used to increase the efficiency of learning described in the previous section. Section~3.1 focuses on clustering semantically similar vocabulary items, whereas Section~3.2 describes preprocessing techniques of grouping in order to obtain stable chunk boundaries.

\subsection{Normalization of Input Representations via Vocabulary Rewriting}
Large-scale Transformer-based language models generally absorb inflectional variation and minor paraphrastic changes through their learned representations, without requiring explicit normalization. In this work, frequently occurring expressions in financial texts are replaced with standardized tags before being fed into the model. This has two main purposes.

\begin{enumerate}
  \item \textbf{Preventing structural errors during circuit construction}\\
  When decomposing a sentence into semantic units to form a circuit, even a slight misalignment at the “connection points” between words or phrases can prevent the circuit from being constructed properly. To avoid this, words with the same grammatical role are normalized to a common tag, ensuring that connections still align correctly after chunking.\\
  \emph{Example:} Frequently appearing nouns such as \texttt{earnings}, \texttt{dividend}, and \texttt{asset price} are mapped to tags representing “financial indicators,” whereas verbs such as \texttt{cuts}, \texttt{raises}, and \texttt{misses} are mapped to tags indicating upward or downward movement. This allows similar vocabulary to be treated as the same component (see Appendix~A for detailed tagging procedures).

  \item \textbf{Handling unknown and rare words}\\
  Here, “unknown and rare words” refer to words that appear at test but were rare or absent in the training data. First, during chunking, syntactic analysis based on CCG assigns each word a type, such as noun type or verb type, allowing the syntactic role of even out-of-vocabulary words to be identified. Furthermore, if a word belongs to one of the financial vocabulary clusters defined during preprocessing, it is mapped to the corresponding tag. If there is no applicable tag, the word is assigned its own rotation-angle parameter, which remains at its initial value. In this way, even when unknown words appear at test, the structure of the quantum circuit remains intact and inference can proceed.
\end{enumerate}
During training, the circuit structure for the entire dataset is fixed in advance, and only the parameters associated with words that appear in the training set are updated. Words that do not appear remain in their initial state, while tagged words are trained with shared parameters, ensuring that the types and circuit structures at inference time are consistent with those used during training. By combining tag sharing with word-specific parameters, the model can learn and infer stably despite the diverse vocabulary characteristic of the financial text.

\subsection{Preprocessing to Preserve Chunk Boundaries}
In our model, sentences are divided into small semantic units called “chunks,” each chunk is converted into a quantum circuit, and a $2\times 2$ density matrix is extracted from each one. But when financial text is converted directly into circuits, differences in word order or variations in spelling and notation often cause connection points to mismatch, which in turn frequently causes circuit construction to fail. To avoid this, we apply rule-based methods that normalize inconsistencies in notation and auxiliary words, thereby stabilizing chunk processing and subsequent circuit construction. Specifically, we apply the following seven rules as preprocessing steps.

\begin{enumerate}
  \item \textbf{Passive constructions and fixed expressions}\\
  Passive constructions such as \texttt{is expected to} and \texttt{was announced to}, as well as fixed expressions frequently used in financial news (e.g., \texttt{is set to}), are treated as single chunks. This standardizes the connection points of verb phrases and stabilizes the links between chunks.

  \item \textbf{Compound prepositions}\\
  Multiword prepositions such as \texttt{due to} and \texttt{as a result of} must be treated as single units; otherwise, their meaning is fragmented. Treating them as single chunks makes the corresponding prepositional phrases well-formed units for composition.

  \item \textbf{Proper nouns such as company names}\\
  We treat expressions such as \texttt{Company A Inc.} and \texttt{Company A} as a single chunk.

  \item \textbf{URLs and symbol sequences}\\
  URLs and special symbol sequences are long character strings that break the assumed syntactic structure for circuit construction. Replacing them with single-word chunks prevents such structural disruptions.

  \item \textbf{\texttt{as + past participle (+ adverb)}}\\
  Subordinate clauses such as \texttt{as expected} and \texttt{as widely} are attached to the subsequent main-clause chunk, preventing these clauses from being split during chunking.

  \item \textbf{\texttt{to + verb}}\\
  Infinitival phrases (\texttt{to} + verb) form separate units if left unmodified. We attach them to their subjects or governing verbs to keep the resulting chunk structure compositionally continuous.

  \item \textbf{Discourse markers at sentence start}\\
  Sentence-initial discourse markers such as \texttt{However,} and \texttt{Therefore,} do not directly contribute to the core semantic structure. If left as independent chunks, they introduce redundant nodes or lead to type mismatches in the circuit. To prevent this, we do not isolate them as separate chunks but instead combine them with the following main clause, thereby avoiding an increase in spurious nodes.
\end{enumerate}

By applying these seven steps, the vocabulary rewriting in Section~3.1 is handled consistently at the chunk level, which stabilizes the construction of quantum circuits.

\subsection{Parameter Sharing}\label{subsec:param-sharing}
The quantum language model used in this study follows the same general framework as QNLP experiments on ion-trap devices~\cite{meichanetzidis2024scalable}, in that it represents each word’s meaning by a circuit with variational parameters. However, with efficiency in mind, we simplify the assignment.

First, each word is assigned a CCG type (such as noun $n$, sentence $s$, or verb $n^r s$) and, at the same time, is mapped via preprocessing rewrite rules to one of “word labels.” For instance, verbs expressing an increase (\textit{increase, rise, climb, improve}, etc.) are grouped under one label, which is distinguished from the label for verbs expressing a decrease. Variational parameters are then assigned to each combination of (post-rewrite word label, CCG type). Therefore,
\begin{itemize}
  \item Even when the CCG type is the same, words that are assigned different labels by the rewrite rules (e.g., the label for rising verbs versus the label for falling verbs) have separate parameter sets.
  \item Conversely, words grouped under the same label (e.g., all “rising verbs”) share a single parameter set corresponding to that type.
  \item Even when the CCG type is the same, words that are not included in any rewrite rule retain their own parameter sets.
\end{itemize}

This design keeps the number of parameters under control while allowing the model to learn distinct quantum states for semantic groups such as increase/decrease or affirmative/negative. Moreover, because word labels are explicitly controlled via rewrite rules, it is structurally clear which groups of words share which parameters, which improves the interpretability of the model.

It should be noted that the CCG compositional rules themselves continue to depend solely on types; the above parameter sharing operates only at the level of lexical semantics. This does not conflict with the type-driven compositional semantic framework of CCG and can be regarded as an implementation-level approximation that abstracts lexical differences into higher-level semantic classes.

\section{Experiment}
In this experiment, we perform 3 class sentiment classification using the Financial PhraseBank data. The Financial PhraseBank contains 4,841 sentences annotated with three sentiment labels (\texttt{positive}, \texttt{neutral}, \texttt{negative}); in this study, we use the 100\%-agreement subset (2,264 sentences). After preprocessing and circuit generation, 2,263 sentences remained. We split these into train/dev/test subsets in a 64\%/16\%/20\% ratio, while preserving the original label distribution (Table~\ref{tab:split}).

\begin{table}[tb]
\centering
\caption{Dataset splits}
\label{tab:split}
\begin{tabular}{lrrrr}
\hline
Split & Total & Class 0 & Class 1 & Class 2 \\
\hline
Train & 1448 & 194 & 889 & 365 \\
Dev & 361 & 48 & 222 & 91 \\
Test & 454 & 61 & 279 & 114 \\
\hline
\end{tabular}
\end{table}

After converting each sentence into a DisCoCat diagram using BobcatParser, we assign an IQP-ansatz quantum circuit of depth $D=4$ using the \texttt{IQPAnsatz} implementation in lambeq. We optimize a class-weighted cross-entropy (CE) loss function, with weights normalized by the inverse class frequencies for the three classes, using Adam with an initial learning rate of $5\times10^{-4}$ and a batch size of 9. We further apply a \texttt{ReduceLROnPlateau} learning-rate scheduler, monitoring dev macro-F1 and halving the learning rate when the score does not improve for two consecutive evaluations (PyTorch settings: factor $0.5$, patience $2$, threshold $2\times 10^{-3}$, cooldown $1$, minimum learning rate $10^{-5}$). Macro-F1 is used as the evaluation metric, and early stopping is applied with a burn-in of 2 epochs and a maximum of 20 epochs. The optimal class-specific thresholds obtained on the development set are $\tau_0{=}0.320$, $\tau_1{=}0.372$, and $\tau_2{=}0.300$.

The results are shown in Table~\ref{tab:results}. Given that the proposed model has approximately $2.2\times10^{4}$ parameters-several orders of magnitude fewer than FinBERT’s 110M parameters-it is unsurprising that its accuracy is lower than that of FinBERT. In addition, because we use a uniform convex combination for density pooling, contextual information spanning multiple chunks is likely to be diluted.

\begin{table}[h]
\centering
\begin{tabular}{lccc}
\toprule
Model & Accuracy & Macro-F1 & Number of Parameters \\
\midrule
QDisCoCirc-inspired (chunked) & 0.696 & 0.551 & $2.2\times10^{4}$ \\
FinBERT& 0.97 & 0.95 & 110M \\
\bottomrule
\end{tabular}
\caption{Comparison of test accuracy on Financial PhraseBank 100\% agree. Results for FinBERT are quoted from \cite{araci2019finbert}.}
\label{tab:results}
\end{table}

\section{Confidence and Sentence-Level Explanation Metrics}
In this section, we introduce two types of scores to enhance the interpretability of the prediction results: (A) probability-based confidence metrics and (B) sentence-level sensitivity metrics based on representation-level interventions. Metrics in group (A) are conventional probability-based scores, whereas metrics in group (B) measure the sensitivity of the output when sentence-level diagram surgery~\cite{Tull2024} is applied to the quantum state representations (Bloch components) of our QDisCoCirc-inspired chunk circuits.

\paragraph{(A) Probability-Based Confidence Metrics (Baseline)}
For the class set $\mathcal{C}=\{0,1,2\}$ and predicted probabilities $p=(p_0,p_1,p_2)$, we define
\[
\textstyle p_{\max}=\max_{c\in\mathcal{C}}p_c,\qquad
\mathrm{prob\_margin}=p_{(1)}-p_{(2)}, \qquad
\mathrm{entropy\_norm}=-\sum_c p_c\log p_c/\log 3,
\]
where $p_{(1)}\ge p_{(2)}\ge p_{(3)}$ denote the largest, second largest, and third largest probabilities, respectively. These metrics capture prediction confidence, but they do not directly indicate which internal factors contributed to a given prediction. 

\paragraph{(B) Metrics Based on In-Model Interventions (Sentence-Level Axis Ablation)}
For the aggregated Bloch vector of the sentence, $r=(r_x,r_y,r_z)$, and the Bloch vector of each label density matrix, $r_c$, we apply a component-wise mask $M\in\{0,1\}^3$:
\[
r[M]=M\odot r,\qquad r_c[M]=M\odot r_c,
\]
where $\odot$ denotes element-wise multiplication. We then recompute the normalized Hilbert--Schmidt similarity as $s_M(c)=s\bigl(\rho(r[M]),\,\rho(r_c[M])\bigr)$. Given $s_M$, we obtain class probabilities $p^M(c)$ via the usual logit-to-softmax mapping, and define
\[
p_{\max}^M=\max_{c} p^M(c).
\]
The masks corresponding to removing, or keeping only, the $z$-axis are
\[
M_{\mathrm{rm},Z}=(1,1,0),\qquad M_{\mathrm{keep},Z}=(0,0,1),
\]
and the corresponding endpoint probabilities are
\[
p_{\mathrm{rm},Z}=p_{\max}^{M_{\mathrm{rm},Z}},\qquad
p_{\mathrm{keep},Z}=p_{\max}^{M_{\mathrm{keep},Z}}
\]
(the $x$- and $y$-axis cases are defined analogously). We then define
\[
\mathrm{comp}_Z=\bigl(p_{\max}-p_{\mathrm{rm},Z}\bigr)_+,\qquad
\mathrm{suff\_gap}_Z=\bigl(p_{\max}-p_{\mathrm{keep},Z}\bigr)_+,\qquad (x)_+=\max(x,0).
\]
Here, $\mathrm{comp}_Z$ quantifies how much the sentence-level prediction confidence drops when the $Z$ component is removed (i.e., the strength of dependence on that axis), whereas $\mathrm{suff\_gap}_Z$ quantifies how much of the original prediction confidence is missing when the prediction is based solely on the $Z$ component (i.e., the insufficiency of that axis alone). For example, a large $\mathrm{comp}_Z$ and a small $\mathrm{suff\_gap}_Z$ can be interpreted as indicating that the overall sentence representation strongly depends on the $Z$ axis and that a similar prediction can be made largely from the $Z$ component alone. Conversely, a large $\mathrm{suff\_gap}_Z$ and a small $\mathrm{comp}_Z$ suggest that the $Z$ component by itself is insufficient and that the prediction relies on a combination with the $X/Y$ components. It is important to emphasize that these are axis-wise sensitivity indicators for the overall sentence representation vector and do not directly specify which words contribute to which axes and to what extent.

Table~\ref{tab:intp} summarizes the relationship between the interpretation metrics and correctness. Here, Cohen’s effect size $d$ is the standardized difference in means between the correctly and incorrectly classified groups; larger $|d|$ indicates that the two distributions are more clearly separated. $r$ is the Pearson correlation and larger $|r|$ indicates a stronger monotonic relationship, i.e., higher scores are more likely to correspond to correct predictions (or vice versa). The metrics that show the strongest differences are $p_{\max}$, $\mathrm{prob\_margin}$, $\mathrm{entropy\_norm}$, $\mathrm{suff\_gap\_x}$, $\mathrm{suff\_gap\_y}$, and $\mathrm{comp\_z}$, all of which show statistically significant differences ($p{<}10^{-13}$).

$\mathrm{comp\_x}$ and $\mathrm{comp\_y}$ also exhibit non-trivial effect sizes, suggesting that sentence-level predictions tend to depend relatively strongly on the contribution of the $Z$ component. At the same time, the large values of $\mathrm{suff\_gap\_x}$ and $\mathrm{suff\_gap\_y}$ indicate that a combination with the $X$ and $Y$ components is also necessary: the model’s decisions are not supported by a single axis, but rather by a combination of axes.

\begin{table}[t]
\centering
\small
\begin{tabular}{lrrrr}
\toprule
Metric & Mean(C) & Mean(I) & Cohen's $d$ & $r$ \\
\midrule
p\_max       & 0.445 & 0.382 &  1.123 &  0.460 \\
prob\_margin & 0.162 & 0.066 &  1.095 &  0.451 \\
entropy\_norm& 0.968 & 0.989 & -1.067 & -0.441 \\
suff\_gap\_x & 0.109 & 0.049 &  1.116 &  0.458 \\
suff\_gap\_y & 0.111 & 0.050 &  1.129 &  0.462 \\
suff\_gap\_z & 0.001 & 0.002 & -0.129 & -0.059 \\
comp\_x      & 0.002 & 0.001 &  0.177 &  0.081 \\
comp\_y      & 0.000 & 0.001 & -0.512 & -0.230 \\
comp\_z      & 0.108 & 0.048 &  1.118 &  0.458 \\
\bottomrule
\end{tabular}
\caption{Relationship between explanation metrics and correctness on the test set ($N{=}454$). Mean(C) and Mean(I) denote the averages for correctly and incorrectly classified instances, respectively. $d$ is Cohen’s effect size and $r$ is the point-biserial correlation.}
\label{tab:intp}
\end{table}

\section{Shallow Transformer Encoder over Bloch-Vector Sequences}
\label{sec:stage2_raw_bloch}
The baseline model represents each sentence as a set of chunk-level Bloch vectors and applies the simple mean of these vectors. This averaging is simple but does not make active use of syntactic cues such as word order or phrase function. In this section, we treat the Bloch-vector sequence $\{(x_t,y_t,z_t)\}_{t=1}^{T}$ as an ordered sequence, attach a CCG type embedding to each chunk, and learn a sentence representation with a small Transformer encoder. This architecture (i) exploits word order and long-range dependencies via positional encodings and self-attention; (ii) injects phrase roles into the attention weights through type embeddings.

The architecture is illustrated in Figure~\ref{fig:bloch-transformer}. Each Bloch vector $r_t\in\mathbb{R}^3$ is first linearly mapped to $\mathbb{R}^{d}$, concatenated with its type embedding $e_t\in\mathbb{R}^{d_{\text{type}}}$, and projected to $h_t\in\mathbb{R}^{d}$. We then add a positional embedding $p_t$ and feed $x_t = h_t + p_t$ into a single-layer, four-head Transformer encoder block ($d{=}128$, feed-forward dimension 256, dropout $0.2$). Masked mean pooling over the encoder outputs yields a sentence vector $\bar h$, which is passed to a three-way linear classifier. We train the model with class-weighted cross-entropy using AdamW (learning rate $10^{-3}$, weight decay $10^{-4}$), and apply a \texttt{ReduceLROnPlateau} scheduler (factor $0.6$, patience $2$) that monitors dev macro-F1; class-wise decision thresholds are tuned on the development set.

As shown in Table~\ref{tab:stage2}, the sequence model improves macro-F1 on the development set by 8.65 points and on the test set by 3.43 points. The gains are particularly pronounced for the minority classes (test class-wise F1: $F1_0{=}0.42$, $F1_1{=}0.83$, $F1_2{=}0.50$).

\begin{sidewaysfigure}[p]
\centering
\resizebox{0.95\textheight}{!}{%
\begin{tikzpicture}[
    >=Latex,
    node distance=7mm and 12mm,
    font=\small,
    box/.style={draw, rounded corners, align=center, inner sep=3pt, minimum height=7mm},
    arrow/.style={->, shorten >=1.5pt}
]

\node[box] (rt) {$r_t\in\mathbb{R}^3$\\(Bloch)};
\node[box, right=of rt] (Wr) {Linear $W_r$\\$\mathbb{R}^3\!\to\!\mathbb{R}^{d_{\text{model}}}$\\($W_r\!\in\!\mathbb{R}^{d_{\text{model}}\times 3}$)};
\node[box, right=of Wr] (ut) {$u_t\in\mathbb{R}^{d_{\text{model}}}$};
\node[box, right=of ut] (concat) {concat $[u_t; e_t]$};
\node[box, right=of concat] (Wc) {Linear $W_c$\\$\mathbb{R}^{\,d_{\text{model}}+d_{\text{type}}}\!\to\!\mathbb{R}^{d_{\text{model}}}$\\($W_c\!\in\!\mathbb{R}^{d_{\text{model}}\times(d_{\text{model}}+d_{\text{type}})}$)};
\node[box, right=of Wc] (ht) {$h_t\in\mathbb{R}^{d_{\text{model}}}$};

\node[box, above=of ht] (pos) {positional embedding\\$p_t\in\mathbb{R}^{d_{\text{model}}}$};

\node[draw,circle, right=6mm of ht, minimum size=3.5mm, inner sep=0pt] (plus) {$+$};
\node[box, right=6mm of plus] (xt) {$x_t=h_t+p_t\in\mathbb{R}^{d_{\text{model}}}$};

\node[box, below=of rt] (ct) {type ID $c_t$\\(CCG)};
\node[box, right=of ct] (Emb) {Embedding table\\$E_{\text{type}}\in\mathbb{R}^{|\mathcal{T}|\times d_{\text{type}}}$};
\node[box, right=of Emb] (et) {$e_t\in\mathbb{R}^{d_{\text{type}}}$};

\draw[arrow] (rt.east)   -- (Wr.west);
\draw[arrow] (Wr.east)   -- (ut.west);
\draw[arrow] (ut.east)   -- (concat.west);
\draw[arrow] (concat.east) -- (Wc.west);
\draw[arrow] (Wc.east)   -- (ht.west);
\draw[arrow] (pos.south) -- (plus.north);
\draw[arrow] (ht.east)   -- (plus.west);
\draw[arrow] (plus.east) -- (xt.west);

\draw[arrow] (ct.east)   -- (Emb.west);
\draw[arrow] (Emb.east)  -- (et.west);
\draw[arrow] (et.east)   -| (concat.south); 

\draw[decorate,decoration={brace,mirror,amplitude=6pt}]
  ($(ct.south west)+(-2pt,-6pt)$) -- ($(et.south east)+(2pt,-6pt)$)
  node[midway, yshift=-12pt] {\scriptsize repeat for $t=1,\dots,T$};

\node[box, below=22mm of xt, minimum width=48mm, minimum height=18mm] (tr) {Transformer Encoder\\
($L{=}1$, heads $=4$, $d_{\text{model}}{=}128$,\\ FFN hidden $=256$, dropout $=0.2$)};

\node[box, right=of tr] (pool) {masked mean pooling\\$\bar{h}\in\mathbb{R}^{d_{\text{model}}}$};

\node[box, right=of pool] (cls) {Linear $\mathbb{R}^{d_{\text{model}}}\!\to\!\mathbb{R}^3$\\softmax};

\draw[arrow] (xt.south) -- node[right] {$x_{1:T}$} (tr.north);
\draw[arrow] (tr.east)  -- node[above] {$z_{1:T}$} (pool.west);
\draw[arrow] (pool.east)-- (cls.west); 

\node[align=left, below=14mm of tr, text width=0.8\linewidth] (note) {
\textbf{Notes:}
$e_t=E_{\text{type}}[c_t]\in\mathbb{R}^{d_{\text{type}}}$.
$W_r$ maps $\mathbb{R}^3\!\to\!\mathbb{R}^{d}$; $W_c$ projects $[u_t;e_t]\!\in\!\mathbb{R}^{d+d_{\text{type}}}$ to $\mathbb{R}^{d}$.
The Transformer output $z_{1:T}$ is pooled with a padding mask to $\bar h$ and fed to a 3-way linear layer with softmax.
};
\end{tikzpicture}%
}
\caption{Architecture of the sequence model that combines Bloch-vector sequences with type embeddings and learns sentence representations using a shallow Transformer encoder.}
\label{fig:bloch-transformer}
\end{sidewaysfigure}

\begin{table}[h]
\centering
\small
\begin{tabular}{lccc}
\toprule
Model & Dev Macro-F1 & Test Acc & Test Macro-F1 \\
\midrule
Baseline (Mean-Pooled Bloch Vectors) & 0.5266 & 0.6960 & 0.5514 \\
Sequence Model (Transformer + Type Embedding) & 0.6131 & 0.6916 & 0.5857\\
\bottomrule
\end{tabular}
\caption{Comparison between the baseline (mean-pooled Bloch-vector representation without order or type information) and the proposed sequence model (Transformer over the Bloch-vector sequence with CCG type embeddings).}
\label{tab:stage2}
\end{table}
We can also track which words assume which types and to what extent each directional component of the Bloch vector $r_t = (r_{x,t}, r_{y,t}, r_{z,t})$ contributes to the decision. 

Let $\overline{\alpha}_t$ denote the attention weight assigned to chunk $t$ (averaged over heads and query positions), $r_t{=}(x_t,y_t,z_t)$ the Bloch vector, and $u\in\mathbb{R}^3$ the unit \emph{readout direction} derived from the final linear layer (or from a class-prototype difference). We define the chunk contribution by
\[
 a_t \;=\; \overline{\alpha}_t\,\langle r_t, u\rangle
 \;=\; \overline{\alpha}_t\,(u_x x_t + u_y y_t + u_z z_t),
\]
and the sentence-level score by $s=\sum_t a_t$. Setting $u{=}(0,0,1)$ recovers the $z$-only view $a_t{=}\overline{\alpha}_t z_t$, but in our experiments $u_x,u_y\!\neq\!0$, indicating that the $x$ and $y$ axes also matter.

For a sentence with $T$ chunks and per-chunk contribution scores $a_t$ defined above, we report
\[
\mathrm{Top20Share}=\frac{\sum_{t\in S}|a_t|}{\sum_{t=1}^{T}|a_t|}
\]
as the fraction of total absolute rationale mass captured by the most influential 20\% of chunks. Attribution statistics show strong sparsity: the top 20\% chunks account for about 58--59\% of the absolute evidential mass (train/dev/test: 0.580/0.585/0.583). The absolute contribution shares on the test split are Bloch (all axes) 40.5\%, CCG-structure 27.2\%, and type-gate 32.2\%.

Furthermore, Table 5 summarizes how the rationale is distributed across three components—semantic, syntactic, and type information—at the sentence level. Their respective contributions are summarized below:
\begin{itemize}
\item Semantic axis (Bloch) contribution: The contribution to prediction made by the Bloch vector components.
\item Syntactic label contribution (CCG): The contribution to prediction made by syntactic information contained in CCG labels themselves (e.g., verb phrase, noun phrase, modifier phrase, etc.).
\item Syntactic gate contribution (type gate): The contribution to prediction made by the weights of the type gate.
\end{itemize}

\begin{table}[t]
\centering
\small
\begin{tabular}{lrrrr}
\toprule
Index (Total Attribution per Channel, Mean/Sentence) & Correct & Incorrect & $\Delta$ (Correct $-$ Incorrect) & Cohen's $d$\\
\midrule
Semantic Axis (Bloch) & 3.30 & 5.11 & $-1.81$ & $-0.65$\\
Syntactic Label (CCG) & 2.16 & 3.56 & $-1.40$ & $-0.76$\\
Syntactic Gate (Type Gate) & 2.67 & 3.95 & $-1.28$ & $-0.56$\\
\bottomrule
\end{tabular}
\caption{Comparison of total attribution per channel on the test set. 
For each sentence $x$, we compute the sum of the absolute values of the attributions 
$a_t^{(q)}$ within the sentence for each channel $q\in\{\mathrm{Bloch},\mathrm{CCG},\mathrm{gate}\}$, 
$\sum_{t\in x}\lvert a_t^{(q)}\rvert$, and show the averages for correct and incorrect sentences.}
\label{tab:channel-mass}
\end{table}

A negative $\Delta$ (correct minus incorrect) indicates that evidence is more dispersed in incorrect sentences. Larger absolute values of Cohen's $d$ similarly indicates a greater gap in total attribution mass between the two groups, reflecting a reliance on more scattered evidence. Empirically, errors increase as evidence becomes thinly distributed across many chunks. Taken together, these results support the intuition that explicitly encoding sentence structure improves both performance and interpretability on real data.

It is worth recalling that the critiques summarized in the introduction~\cite{jain2019attention,wiegreffe2019attention}, question the practice of equating visualized attention weights with explanations. In our model, attention is not used as a visualization but as a mechanistic variable that regulates the flow of information through the type channels. What is explained, therefore, is not merely “which words were highlighted” in an attention map; rather, attention is interpreted as a routing mechanism for type-level information. By analyzing causal contributions along both structural (CCG) and mechanistic (type-gate) axes, we obtain a more principled form of explainability.

Beyond observational (correlational) evaluation, the model can support validity checks based on interventional manipulations of the input. Let the model score (for example, a logit) be $f_\theta(x)\in\mathbb{R}$, and let $a_i(x)\in\mathbb{R}$ denote the explanation weight for the $i$-th component (e.g., token) of input $x$. Define $S\subseteq\{1,\dots,|x|\}$ as the set of manipulated components and $\delta\in\mathbb{R}$ as the intervention strength. We write $\mathcal{I}_{S,\delta}(x)$ for the intervention operator that modifies only the elements in $S$ of $x$ (covering task-specific instantiations such as word polarity reversal, scaling of numerical quantities, and span permutation).

Then we can evaluate the causal relationship between explanation weights and predictions. Conceptually, this belongs to the same family as existing intervention principles, but we instantiate it through three complementary metrics: (i) Directional Consistency (DC), (ii) Proportional Response (PR), and (iii) Monotonicity Violation Rate (MVR).

\medskip

\noindent
\noindent\textbf{(i) Directional Consistency Rate (Directional Consistency; DC)}:
\begin{equation}
\mathrm{DC}
=\mathbb{E}_{x,S,\delta}\Big[
\mathbf{1}\Big\{
\mathrm{sign}\!\Big(\sum_{i\in S} a_i(x)\,\delta\Big)
=
\mathrm{sign}\!\Big(f_\theta(\mathcal{I}_{S,\delta}(x)) - f_\theta(x)\Big)
\Big\}
\Big].
\label{eq:dc}
\end{equation}
This metric measures whether the model output changes in the direction indicated by the aggregated intervention$\sum_{i\in S} a_i(x)\,\delta$.

\medskip
\noindent\textbf{(ii) Proportional Response Coefficient (Proportional Response; PR)}:
\begin{equation}
\mathrm{PR}
=\mathrm{Corr}\!\left(
\sum_{i\in S} a_i(x)\,\delta,\;
f_\theta(\mathcal{I}_{S,\delta}(x)) - f_\theta(x)
\right),
\label{eq:pr}
\end{equation}
This metric captures the linear relationship between intervention strength (weighted by attribution) and the resulting change in model output.

\medskip
\noindent\textbf{(iii) Monotonicity Violation Rate (MVR):}  
Given a sequence $\delta_1<\delta_2<\cdots<\delta_K$,  
\begin{equation}
\mathrm{MVR}
=\mathbb{E}_{x,S}\!\left[
\frac{1}{K-1}\sum_{k=1}^{K-1}
\mathbf{1}\!\left\{
\Big(f_\theta(\mathcal{I}_{S,\delta_{k+1}}(x)) - f_\theta(x)\Big)
<
\Big(f_\theta(\mathcal{I}_{S,\delta_k}(x)) - f_\theta(x)\Big)
\cdot \mathrm{sign}\!\Big(\sum_{i\in S} a_i(x)\Big)
\right\}
\right],
\label{eq:mvr}
\end{equation}
measures local violations of monotonicity in the expected direction as the intervention strength $\delta_k$ increases.

\medskip

\noindent
Conventional observational fidelity metrics, such as deletion/insertion curves, primarily measure correlations between output and explanation. In contrast, our framework evaluates the operational validity of explanations by explicitly designing input interventions to elicit counterfactual responses, and by jointly requiring directional consistency, proportional response, and monotonicity through (\ref{eq:dc})–(\ref{eq:mvr}).

Furthermore, while the \emph{no-influence}, \emph{diagram surgery}, and \emph{rewrite explanations} proposed by Tull \emph{et al.}~\cite{Tull2024} are conceptual tests based on schematic diagrams and do not define quantitative metrics, DC, PR, and MVR function as operationally formalized counterparts. Diagram surgery corresponds to severing or rewiring partial circuits; in our setting, this is quantified by measuring DC, PR, and MVR under interventions such as zeroing out type gates, applying attention masks, or perturbing the Bloch direction \(u\).

No-influence serves as a test for zero influence: when the corresponding pathway is blocked by a gate, the output difference vanishes, corresponding to PR$\to 0$ and MVR$\to 0$. Finally, rewrite explanations verify invariance under equivalence-preserving transformations $D\!\to\!D'$, where we require the rewrite discrepancy $\Delta_{\text{rewrite}} = \lVert f_\theta(D)-f_\theta(D') \rVert$ to remain small; in such cases, DC, PR, and MVR are expected to be negligible.

\section{Conclusion}
This study evaluated a \emph{QDisCoCirc}-inspired, chunked diagram-to-circuit QNLP model for three-class sentiment classification of financial texts in a classical-simulation setting. By augmenting QDisCoCirc with a shallow Transformer encoder that models sentence structure, we showed that it is possible to partially overcome the limitations of mean pooling. Moving forward, it will be important to balance performance and scalability by exploiting the unique features of QDisCoCirc, developing new compositional rules that combine information between chunks at the quantum level, and incorporating circuit-compression techniques such as XZ reduction. Implementing and evaluating financial sentiment analysis tasks on actual devices, such as superconducting quantum processors, is another key next step.

While this paper focused on sentence-level sentiment classification, future challenges include extending the approach to tasks involving multi-sentence reasoning, such as FinQA~\cite{chen2021finqa} and ConvFinQA~\cite{chen2022convfinqa}. The FinBen repository organizes 36 datasets and 24 tasks into seven broad categories for evaluating the performance of large language models in the financial domain. From the perspective of classification versus reasoning, these categories can be organized as shown in the table below.

\vspace{0.3em}
\begin{tabular}{p{0.25\linewidth}p{0.45\linewidth}p{0.23\linewidth}}
\toprule
Category & Representative Task Examples & Problem Type \\\midrule
Information Extraction (IE) & NER, Relation Extraction & Multiclass / Multilabel Classification \\
Text Analysis (TA) & Sentiment Analysis, Financial NLI & Multiclass Classification \\
Question Answering (QA) & Multiple-choice QA & Multiclass Classification \\
Risk Management (RM) & Credit Scoring, Fraud Detection & Binary / Multiclass Classification \\\midrule
Text Generation (TG) & Headline Generation & Generation + Reasoning \\
Forecasting & Stock Price Time-series Prediction & Regression + Reasoning \\
Decision-making Support (DM) & Portfolio Optimization & Search / Optimization + Reasoning \\
\bottomrule
\end{tabular}

\vspace{0.5em}

Of the seven categories listed above, the four categories IE, TA, QA, and RM can be reduced to multi-class or multi-label classification, making it possible to apply the model discussed in this paper. Our quantum compositional model can in principle cover these four task categories. The remaining three task domains-TG, Forecasting, and DM-involve inference problems that require generation, regression, or optimization. These tasks will require:
\begin{enumerate}
  \item long-range-dependency circuits that cover the entire inference chain,
  \item quantum computation subnetworks that handle numerical operations, and
  \item mechanisms for retaining state between inference steps.
\end{enumerate}

Although constraints on circuit depth and shot count remain, it is possible on actual devices to construct quantum circuits that handle longer text chunks. For each chunk, the expectation values of the Pauli $X/Y/Z$ operators can be estimated from shot measurements as
\[
\mathbf{b}^{(c)} = \bigl(\langle \sigma_x \rangle^{(c)},\, \langle \sigma_y \rangle^{(c)},\, \langle \sigma_z \rangle^{(c)}\bigr),
\]
which are then stored as the Bloch vector for that chunk. The resulting sequence of vectors can be used as input to a classical inference module (e.g., a Transformer). Training can be completed on the classical side, and only during inference are the parameterized circuits executed on the physical device. Incorporating dynamic circuits together with measurement-error mitigation and probabilistic error-reduction techniques improves both shot efficiency and robustness. As these methods mature, the range of inputs that can be handled on device is expected to increase.

\section*{Declaration of generative AI and AI-assisted technologies}
During the preparation of this manuscript, the author used generative AI and AI-assisted tools to assist with manuscript drafting, language refinement, code drafting and debugging, and discussion of the research. The author directed the project, revised and verified the manuscript and code, and takes full responsibility for the final manuscript.

\section*{Acknowledgements}
We thank Konstantinos Meichanetzidis for helpful discussions and feedback.

\newpage

\section*{Appendix A. Rewrite Rules Specialized for Financial Text}
Before mapping sentences to quantum circuits, expressions commonly used in financial text are replaced with semantically explicit tags. This ensures stable connections between words and phrases and prevents the circuit structure from collapsing. There are seven rules, which are grouped into three levels: lexical, phrasal, and syntactic.

\paragraph{1. Lexical Level: Normalizing Words}
\begin{enumerate}
  \item \textbf{Unifying copulas (be-verbs)}\\
  Map \texttt{is}, \texttt{are}, \texttt{was}, etc.\ to a single tag.
  \item \textbf{Normalization of numbers and units}\\
  Normalize terms such as \texttt{million}, \texttt{percent}, \texttt{USD}, and \texttt{kWh} into numeric/unit tags.
  \item \textbf{Clustering upward/downward movements}\\
  Group upward-movement verbs such as \texttt{rise}, \texttt{increase}, \texttt{surge} and downward-movement verbs such as \texttt{fall}, \texttt{decline} into corresponding tags.
\end{enumerate}

\paragraph{2. Phrase Level: Normalizing Meaning}
\begin{enumerate}\setcounter{enumi}{3}
  \item \textbf{Tagging prepositional meanings}\\
  Tag prepositions according to their semantic roles, e.g., ``\texttt{in}'' $\to$ \texttt{location\_in}, ``\texttt{by}'' $\to$ \texttt{agent\_by}, following a consistent role-tagging scheme.
  \item \textbf{Tagging finance-specific relative pronouns}\\
  Tag words such as ``\texttt{that}'', ``\texttt{which}'', and ``\texttt{where}'' in a way that reflects their use in financial news.
  \item \textbf{Combining compound prepositions}\\
  Merge multiword prepositional phrases such as ``\texttt{due to}'' and ``\texttt{as a result of}'' into unified tags.
\end{enumerate}

\paragraph{3. Syntactic Level: Normalizing Function Words}
\begin{enumerate}\setcounter{enumi}{6}
  \item \textbf{Unifying comparative expressions}\\
  Standardize words and phrases such as ``\texttt{higher}'', ``\texttt{lower}'', ``\texttt{better}'', ``\texttt{worse}'', and ``\texttt{compared}''.
\end{enumerate}

To prevent interference between rules, the rewrites are applied in the order ``lexical $\to$ phrase $\to$ syntax.'' First, orthographic and lexical variations are eliminated at the word level; next, semantic blocks are stabilized at the phrase level; and finally, the syntactic roles of the entire sentence are organized. Within the same level, the rules are designed to minimize overlap in their vocabulary and patterns. 


\newpage

\begin{thebibliography}{99}

\bibitem{araci2019finbert}
Araci, D. (2019).
FinBERT: Financial sentiment analysis with pre-trained language models.
arXiv:1908.10063.

\bibitem{belinkov2019analysis}
Belinkov, Y., Glass, J. (2019).
Analysis methods in neural language processing: a survey.
\emph{Transactions of the Association for Computational Linguistics}, \textbf{7}, 49--72.

\bibitem{chen2021finqa}
Chen, Z. et al. (2021).
FinQA: A dataset of numerical reasoning over financial data.
arXiv:2109.00122.

\bibitem{chen2022convfinqa}
Chen, Z. et al. (2022).
ConvFinQA: Exploring the chain of numerical reasoning in conversational finance question answering.
arXiv:2210.03849.

\bibitem{meichanetzidis2024scalable}
Duneau, T., Bruhn, S., Matos, G., Laakkonen, T., Saiti, K., Pearson, A.,
Meichanetzidis, K., Coecke, B. (2024).
Scalable and interpretable quantum natural language processing:
an implementation on trapped ions.
arXiv:2409.08777.

\bibitem{fong2017interpretable}
Fong, R.\,C., Vedaldi, A. (2017).
Interpretable explanations of black boxes by meaningful perturbation.
In \emph{Proceedings of the IEEE International Conference on Computer Vision (ICCV)}, 3449--3457.



\bibitem{hewitt2019structural}
Hewitt, J., Manning, C.\,D. (2019).
A structural probe for finding syntax in word representations.
In \emph{Proceedings of the 2019 Conference of the North American Chapter of the Association for Computational Linguistics:
Human Language Technologies}, 4129--4138.

\bibitem{jain2019attention}
Jain, S., Wallace, B.\,C. (2019).
Attention is not explanation.
In \emph{Proceedings of the 2019 Conference of the North American Chapter of the Association for Computational Linguistics:
Human Language Technologies}, 3543--3556.

\bibitem{laakkonen2024algorithms}
Laakkonen, T., Meichanetzidis, K., Coecke, B. (2024).
Quantum algorithms for compositional text processing.
arXiv:2408.06061.

\bibitem{patterson2021carbon}
Patterson, D.\,A., Gonzalez, J., Le, Q.\,V., Liang, C., Munguia, L.-M.,
Rothchild, D., So, D.\,R., Texier, M., Dean, J. (2021).
Carbon emissions and large neural network training.
arXiv:2104.10350.

\bibitem{rudin2019stop}
Rudin, C. (2019).
Stop explaining black box machine learning models for high stakes decisions and use interpretable models instead.
\emph{Nature Machine Intelligence}, \textbf{1}, 206--215.

\bibitem{Tull2024}
Sean Tull, Robin Lorenz, Stephen Clark, Ilyas Khan, and Bob Coecke.
\newblock Towards Compositional Interpretability for XAI.
\newblock arXiv:2406.17583, 2024.

\bibitem{simonyan2014deep}
Simonyan, K., Vedaldi, A., Zisserman, A. (2014).
Deep inside convolutional networks: visualising image classification models and saliency maps.
In \emph{Workshop at International Conference on Learning Representations}.

\bibitem{strubell2019energy}
Strubell, E., Ganesh, A., McCallum, A. (2019).
Energy and policy considerations for deep learning in NLP.
In \emph{Proceedings of the 57th Annual Meeting of the Association for Computational Linguistics}, 3645--3650.

\bibitem{tatsat2025beyond}
Tatsat, H., Shater, A. (2025).
Beyond the black box: interpretability of LLMs in finance.
arXiv:2505.24650.

\bibitem{tian2023ccg-easa}
Tian, Y., Chen, W., Hu, B., Song, Y., Xia, F. (2023).
End-to-end Aspect-based Sentiment Analysis with Combinatory Categorial Grammar.
\textit{Findings of the Association for Computational Linguistics: ACL 2023}, 13597--13609.

\bibitem{wiegreffe2019attention}
Wiegreffe, S., Pinter, Y. (2019).
Attention is not not explanation.
In \emph{Proceedings of the 2019 Conference on Empirical Methods in Natural Language Processing and
the 9th International Joint Conference on Natural Language Processing}, 11--20.

\bibitem{wu2023bloomberggpt}
Wu, S., Irsoy, O., Lu, S., Dabravolski, V., Dredze, M., Gehrmann, S.,
Kambadur, P., Rosenberg, D., Mann, G. (2023).
BloombergGPT: A large language model for finance.
arXiv:2303.17564.

\bibitem{yang2023fingpt}
Yang, H., Liu, X.-Y., Wang, C.\,D. (2023).
FinGPT: Open-Source Financial Large Language Models.
arXiv:2306.06031.

\bibitem{zhao2024llm-supertagger}
Zhao, J., Penn, G. (2024).
LLM-supertagger: Categorial Grammar Supertagging via Large Language Models.
\textit{Findings of the Association for Computational Linguistics: EMNLP 2024}, 697--705.
\end{thebibliography}
\end{document}